\documentclass[12pt]{iopart}

\usepackage{graphicx}

\usepackage{caption}
\usepackage{subcaption}

\usepackage{adjustbox}
\usepackage{aas_macros}

\usepackage{hyperref}
\usepackage{bm}
\usepackage{color}
\usepackage{amssymb}
\usepackage{breqn}

\def\aligned{\vcenter\bgroup\let\\\cr
\halign\bgroup&\hfil${}##{}$&${}##{}$\hfil\cr}
\def\endaligned{\crcr\egroup\egroup}

\begin{document}

\title[Interpreting a gravitational-wave burst]{Interpreting gravitational-wave burst detections: constraining source properties without astrophysical models}

\author{Bence B\'ecsy$^1$, Peter Raffai$^{1,2}$, Kiranjyot Gill$^3$, Tyson B.~Littenberg$^4$, Margaret Millhouse$^5$, Marek J.~Szczepa\'nczyk$^6$}

\address{$^1$ Institute of Physics, E\"otv\"os University, 1117 Budapest, Hungary}
\address{$^2$ MTA-ELTE Extragalactic Astrophysics Research Group, 1117 Budapest, Hungary}
\address{$^3$ Harvard-Smithsonian Center for Astrophysics, 60 Garden Street, Cambridge, MA 02138, USA}
\address{$^4$ NASA Marshall Space Flight Center, Huntsville, AL 35812, USA}
\address{$^5$ OzGrav, University of Melbourne, Parkville, Victoria 3010, Australia}
\address{$^6$ University of Florida, Gainesville, FL 32611, USA}

\ead{becsybence@caesar.elte.hu}
\vspace{10pt}
\begin{indented}
\item[]March 2020
\end{indented}

\begin{abstract}
We show that for detections of gravitational-wave transients, constraints can be given on physical parameters of the source without using any specific astrophysical models. Relying only on fundamental principles of general relativity, we can set upper limits on the size, mass, and distance of the source solely from characteristics of the observed waveform.
If the distance of the source is known from independent (e.g.~electromagnetic) observations, we can also set lower limits on the mass and size.
As a demonstration, we tested these constraints on binary black hole signals observed by the LIGO and Virgo detectors during their first and second observing runs, as well as on simulated binary black hole and core-collapse supernova signals reconstructed from simulated detector data.
We have found that our constraints are valid for all analyzed source types, but their efficiency (namely, how far they are from the true parameter values) strongly depends on the source type, ranging from being in the same order of magnitude to a several orders of magnitude difference.
In cases when a gravitational-wave signal is reconstructed without waveform templates and no astrophysical model on the source is available, these constraints provide the only quantitative characterization of the source that can guide the astrophysical modeling process.
\end{abstract}

%
% Uncomment for keywords
% \vspace{2pc}
% \noindent{\it Keywords}: XXXXXX, YYYYYYYY, ZZZZZZZZZ
%
% Uncomment for Submitted to journal title message
%\submitto{\JPA}
%
% Uncomment if a separate title page is required
%\maketitle
% 
% For two-column output uncomment the next line and choose [10pt] rather than [12pt] in the \documentclass declaration
%\ioptwocol
%

\section{Introduction} \label{sec:intro}
\begin{sloppypar}
The two detectors of the Laser Interferometer Gravitational-Wave Observatory (LIGO, \cite{aligo}) have finished two observing runs so far (designated O1 and O2), during which they observed gravitational waves (GWs) from coalescences of binary black holes (BBHs, see e.g.~\cite{O1BBH, O2BBH}) and a binary neutron star \cite{bns}. The Virgo detector \cite{avirgo} was also in observing mode during the last month of O2 (August 2017), and contributed to the observation of additional GW events \cite{gw170814, bns}. After the end of O2 on 25 August 2017, all three detectors had been under commissioning break, and started the third observing run (O3) in April 2019. In the near future, the KAGRA detector \cite{kagra} will also join the network of advanced GW detectors \cite{prospects}.
\end{sloppypar}

Besides compact binary coalescences (CBC), for which template-based searches are available, the LIGO-Virgo Collaboration (LVC) also carries out searches for generic GW transients (a.k.a.~\emph{bursts}), where no accurate waveform models exist (see e.g.~\cite{O1_allsky}). Although all GW transients that have been detected so far were identified as signals from CBC sources \cite{O1_allsky, O2_allsky}, the chance of detecting a GW burst without an associated source model is increasing with the space-time volume surveyed by the expanding and improving network of GW detectors. The current expectation is that by 2024, the two LIGO detectors, the Virgo detector, and the KAGRA detector will all operate at their design sensitivities \cite{prospects}.

The LVC employs multiple algorithms to detect burst signals by looking for statistically significant coherent excess power in all detectors' datastreams; and also to reconstruct their waveforms using generic base functions \cite{bayeswave, cwb, BWPE, olib}.
To determine properties of the GW source\footnote{Here and throughout the paper we use the term ``GW source'' in a restricted sense, referring only to the part of an astrophysical object that emits an observable GW. For example, in a core-collapse supernova, the term ``GW source'' refers only to the GW-emitting core, and not to the entire exploding star.}, one needs to extract information from the reconstructed GW waveform, which carries information about the dynamics of the source. The quadrupole formula describing an emitted GW (see e.g.~equations (3.67) and (3.68) in \cite{maggiore_book}) is not invertible, and thus the dynamics of the source cannot be reconstructed entirely from the observed waveform. One can assume a specific source model to overcome this problem, and find the best-fit values of model parameters. This method, however, cannot be used in case of the discovery of a new, unexpected source type, where no model predicting the GW waveform is available.

In this paper, for the first time, we address the problem of extracting astrophysical information about the source from the observed GW waveform without assuming any specific source model. Using the properties of the GW waveform reconstructed model-independently, we can set upper limits on the characteristic size, the characteristic mass and the luminosity distance of the source. If the source distance is known (e.g.~from electromagnetic observations), we can also set lower limits on the characteristic size and mass. These constraints utilize fundamental principles of general relativity, which has been extensively tested and confirmed (for a review, see e.g.~\cite{testsofGR}).  The constraints can be calculated for any transient GW signal, and for GW bursts without a clear model on their astrophysical source, they provide the only quantitative characterization of the GW source. Note that most of the formulae describing our constraints are similar to and inspired by formulae derived before the first detection of GWs to give order of magnitude predictions of the expected properties of GWs using educated guesses on source parameters (see e.g.~\cite{thorne_review, hughes_review, schutz_book}), however we propose to use them ``backwards'', i.e.~constraining parameters of GW sources using actual measured properties of their detected and reconstructed signals.

To test the validity and efficiency of our proposed constraints, we applied them on both observed and simulated GW signals: i) five BBH signals detected during O1 and O2; ii) hundreds of simulated BBH signals; and iii) tens of simulated core-collapse supernova (CCSN) signals. We embedded all simulated signals into simulated Gaussian noise to create realistic samples of data we can feed as an input to our waveform reconstruction algoritm. To reconstruct the waveforms of both simulated and observed signals, we used BayesWave, one of the algorithms used by the LVC, which applies Bayesian methods to estimate parameters and reconstruct waveforms of GW bursts (see \cite{bayeswave} and \cite{chirplet} for details).

This paper is organized as follows. We introduce our constraints in Section \ref{sec:derivation}. We present tests of the constraints with observed and simulated signals in Section \ref{sec:testing}, and discuss some practical aspects of their application in Section \ref{sec:discussion}. We derive conclusions and discuss potential implications of our work in Section \ref{sec:conclusion}. \ref{sec:bw_appendix} provides additional information on how the signal parameters used to calculate our constraints can be estimated from the data.

\section{Description of constraints}
\label{sec:derivation}

\begin{sloppypar}
In this section, we introduce different model-independent constraints on astrophysical parameters of GW sources. We describe upper limits we can set on the characteristic size (see Section \ref{ssec:size}), characteristic mass (see Section \ref{ssec:mass}), and luminosity distance (see Section \ref{ssec:distance}) of the source based solely on the observed GW waveform. In Section \ref{ssec:mass_known_distance}, we investigate the scenario when the source distance is known from non-GW observations, and we describe the resulting lower limits on the characteristic size and mass of the source.
\end{sloppypar}

\subsection{Upper limit on the characteristic size}
\label{ssec:size}

We can set an upper limit on the characteristic size ($D$) of a GW source (see footnote 1 on page 2) using the central frequency of the observed GW signal, $f_0^{\rm GW}$ (see \ref{sec:bw_appendix}). Based on the quadrupole nature of the dominant term in GW emission (see e.g.~Chapter 3.3.5 in \cite{maggiore_book} for details), we can assume that:
\begin{equation}
f_0^{\rm GW} \simeq 2 f^{\rm s},
\label{eq:assumption}
\end{equation}
where $f^{\rm s}$ is the frequency of the motion generating the GW. (\ref{eq:assumption}) becomes an exact equivalence if the motion is a harmonic oscillation at non-relativistic speeds. Causality requires that the typical velocity in the source, $v$, must at all time be less than the speed of light, $c$, i.e.~$c>v \equiv \pi f^{\rm s} D$, where $D$ is defined as the characteristic size (and $D/2$ as the characteristic \emph{radius}) of the GW source. Combining this criterion with (\ref{eq:assumption}) gives the following upper limit on $D$:
\begin{equation}
D_{\rm max} = \frac{2 c}{\pi} \frac{1}{f_0^{\rm GW}} \simeq 1909 \left( \frac{f_0^{\rm GW}}{100 \ \mathrm{Hz}} \right)^{-1}  \ \mathrm{km}.
\label{eq:size_limit}
\end{equation}

It has been shown that $f_0^{\rm GW}$ can be estimated robustly for a wide range of signal morphologies with a median statistical error of $\sim$10\% of the signal bandwidth (see \cite{BWPE}). Thus determining $D_{\rm max}$, which only depends on $f_0^{\rm GW}$, is straightforward for any detected transient GW signal.

In most cases the cosmological redshift ($z$) of the source is unknown. It has been neglected in (\ref{eq:size_limit}), because taking it into account would introduce a $(1+z)^{-1}$ factor in the expression of $D_{\rm max}$, which would reduce the $D_{\rm max}$ value we get in (\ref{eq:size_limit}). Thus, the way we define $D_{\rm max}$ in (\ref{eq:size_limit}) remains a valid upper limit for all sources, regardless of their $z$\footnote{We neglected peculiar velocities here, which, in principle, could artificially decrease our upper limit. However, we do not expect peculiar motions to overcome the effect of cosmological redshift for sources in cosmological distances.}. Alternatively, we can say that with (\ref{eq:size_limit}), we can set an upper limit on the characteristic size measured in the detector frame, which is $(1+z)$ times larger than in the source frame.

It should be noted, that there is an inescapable limitation of this constraint: it can only give information about the GW source, but not necessarily about the entire astrophysical object. E.g.~for an eccentric BBH system in the phase of repeated bursts (see e.g.~\cite{kocsis_ebbh_rb}), we would only observe GWs at pericenter passages. This means that the upper limit on the characteristic size obtained with (\ref{eq:size_limit}) would only be an upper limit on the pericenter distance, but not on the semi-major axis or on the apocenter distance of the system.

\subsection{Upper limit on the characteristic mass}
\label{ssec:mass}
Knowing the previously described upper limit on $D$, we can also set an upper limit on the GW source's (see footnote on page 2) characteristic mass, $M$. This is based on the fact that there is an absolute upper bound on the compactness of the GW source:
\begin{equation}
\frac{M}{R} \leq \frac{c^2}{G},
\label{eq:compactness_limit}
\end{equation}
where $R=D/2$ is the characteristic radius of the source, $G$ is the gravitational constant, and the equality corresponds to a maximally spinning Kerr black hole. Combining this with (\ref{eq:size_limit}), we get the following upper limit on $M$:
\begin{equation}
M_{\rm max} = \frac{c^3}{\pi G} \frac{1}{f_0^{\rm GW}} \simeq 646 \left( \frac{f_0^{\rm GW}}{100 \ \mathrm{Hz}} \right)^{-1} \ M_{\odot},
\label{eq:mass_limit}
\end{equation}
where $M_{\odot}$ is the mass of the Sun.

Similarly to the case of $D_{\rm max}$, it is straightforward to calculate $M_{\rm max}$ for any observed transient GW signal, and it gives a valid upper limit on $M$ regardless of the $z$ of the source. Also similarly to $D_{\rm max}$, $M_{\rm max}$ is only a valid upper limit on the GW source, but not necessarily on the entire astrophysical object. E.g.~in case of a CCSN, $M_{\rm max}$ constrains the mass of the GW-emitting core, but not the entire exploding star. For more details see the discussion at the end of Section \ref{ssec:size}.

\subsection{Upper limit on the luminosity distance}
\label{ssec:distance}
General relativity implies an upper limit on the luminosity of any emission process regardless of the mass of the source, which comes from the fact that a higher luminosity would require such a high energy density in the source, that it would collapse into a black hole (see e.g.~Chapter 9.4 of \cite{schutz_book} for details):
\begin{equation}
L_{\rm max} = \frac{c^5}{G} \simeq 3.6 \times 10^{52} \ \mathrm{W}.
\label{eq:lumlimit}
\end{equation}

We can set an upper limit on the luminosity distance of a GW source (see footnote 1 on page 2), $d_L$, by comparing the observed GW luminosity, $L_{\rm GW}$, to $L_{\rm max}$. For a GW signal, assuming isotropic emission, we can calculate the time-dependent luminosity of the source using the following formula (see e.g.~equation (17) in \cite{schutz_lrr}):
\begin{equation}
L_{\rm GW} (\bar{t}) = \frac{c^3}{4 G} d_L^2 \left( \dot{h} (t) \right)^2,
\label{eq:gw_lum}
\end{equation}
where $\dot{h} (t)$ is the time derivative of the detected waveform, $h (t)$, and $\bar{t}$ and $t$ are measured in the source-frame and the detector-frame, respectively.

Combining (\ref{eq:lumlimit}) with (\ref{eq:gw_lum}), and neglecting the cosmological Doppler-effect (i.e.~assuming that $\bar{t} -t$ is constant), we get an upper limit on the luminosity distance of the source:
\begin{equation}
d_{L, {\rm max}} = 2 c \left( \dot{h}^{*} \right)^{-1} \simeq 19.4  \left( \frac{\dot{h}^{*}}{10^{-18} \ \mathrm{s^{-1}}} \right)^{-1} \ \mathrm{Gpc},
\label{eq:distanceupper1}
\end{equation}
where $\dot{h}^{*}$ is the maximum of $\dot{h} (t)$ throughout the observed signal (see \ref{sec:bw_appendix}). Note that a similar analysis can give a distance estimate specifically for BBHs \cite{basic_gw150914}.

The cosmological Doppler-effect can be taken into account by introducing a $(1+z)^{-1}$ factor in (\ref{eq:distanceupper1}), which gives the following equation:
\begin{equation}
d_{L, {\rm max}} = 2 c \frac{1}{\dot{h}^{*} (1+z_{\rm max})},
\label{eq:zcorr}
\end{equation}
% = (1+z_{\rm max}) \frac{c}{H_0} \int_0^{z_{\rm max}} \frac{{\rm d} z'}{E(z')}
where $z_{\rm max}$ is the redshift that corresponds to a luminosity distance of $d_{L, {\rm max}}$. As $z_{\rm max}$ depends on $d_{L, {\rm max}}$, (\ref{eq:zcorr}) cannot be solved analytically. The dependence of $z_{\rm max}$ on $d_{L, {\rm max}}$ is set by the cosmological model, for which we assume a flat $\Lambda$CDM cosmology with parameters reported in the ``Planck + WP + highL + BAO'' column of Table 5 in \cite{Planck2013}. The self-consistent solution of (\ref{eq:zcorr}) can be found by iteration (see Section \ref{ssec:results}), which we have found to be converging in all cases, and which significantly lowers the upper limit given by (\ref{eq:distanceupper1}).

\subsection{Lower limits for sources with known distances}
\label{ssec:mass_known_distance}

In this section we examine what constraints we can set on a GW source (see footnote on page 2) for which we know its comoving distance, $d_C$, e.g.~from identification of its host galaxy (as it happened for GW170817, see \cite{mma}).

The $+$ and $\times$ polarizations of the emitted GW are given by the quadrupole formula\footnote{We follow the notations used in Section 3.3 of \cite{maggiore_book}}:
\begin{equation}
h_+ (t) = \frac{G}{c^4 d_C} \left( \ddot{M}_{11} - \ddot{M}_{22} \right),
\end{equation}

\begin{equation}
h_{\times} (t) = \frac{2 G}{c^4 d_C} \ddot{M}_{12},
\end{equation}
where $M_{ij}$ are the second momenta of mass in the source frame where the observer is at the positive $z$ direction, and the two dots denote the second time derivative. The second time derivative of all $M_{ij}$ components must in all cases satisfy:
\begin{equation}
\ddot{M}_{ij} \leq M c^2,
\label{eq:ddot_M_limit}
\end{equation}
where $M$ is the characteristic mass of the source. From (\ref{eq:ddot_M_limit}), we can set an upper limit on the GW amplitudes:
\begin{equation}
h_{+,\times} (t) \lesssim \frac{2 G M}{d_C c^2},
\end{equation}
which is valid for all $t$.

Combining $h_+$ and $h_{\times}$ in the output of a GW detector implies the following upper limit on the maximum amplitude value of the observed $h(t)$ GW signal (see \ref{sec:bw_appendix}):

\begin{equation}
h^{*} \lesssim \frac{4 G M}{d_C c^2}.
\end{equation}

Thus by knowing the comoving distance, we can set a lower limit on the GW source's (see footnote on page 2) characteristic mass:
\begin{equation}
M_{\rm min} = \frac{c^2}{4 G} h^{*} d_C \simeq 0.52 \left( \frac{h^{*}}{10^{-21}} \right) \left( \frac{d_C}{100 \ \mathrm{Mpc}} \right) \ M_{\odot},
\label{eq:mass_lower}
\end{equation}
Using (\ref{eq:compactness_limit}), we can follow a similar argument to the one we used in Section \ref{ssec:mass}, and we can convert $M_{\rm min}$ into the following lower limit on the characteristic size of the GW source:
\begin{equation}
D_{\rm min} = \frac{1}{2} h^{*} d_C \simeq 1.54 \left( \frac{h^{*}}{10^{-21}} \right) \left( \frac{d_C}{100 \ \mathrm{Mpc}} \right) \ \mathrm{km},
\label{eq:size_lower}
\end{equation}

The cosmological Doppler-effect does not affect $h^{*}$, so there is no need for a redshift correction in (\ref{eq:mass_lower}) and (\ref{eq:size_lower}). 

Note that when the source distance is known, we can also refine our upper limits on the characteristic size and mass by including a $(1+z)^{-1}$ factor in the expressions of $D_{\rm max}$ and $M_{\rm max}$ (see (\ref{eq:size_limit}) and (\ref{eq:mass_limit})).

%\begin{comment}
%In Table \ref{tab:limits} we show the summary of the upper limits derived in this section.
%\end{comment}

\section{Testing the constraints}
\label{sec:testing}

To test the constraints introduced in Section \ref{sec:derivation}, we calculated them for both observed and simulated GW signals, and compared them to the reference values of the constrained parameters ($M_{\rm ref}$, $D_{\rm ref}$, $d_{L, \mathrm{ref}}$). We take these reference values to be the point estimates given by the model-based analysis for observed signals; and the preset parameter values of the simulations for simulated signals. We describe the signals we used and the methods we applied for these tests in Section \ref{ssec:testing_methods}, and present our results in Section \ref{ssec:results}.

\subsection{Methods}
\label{ssec:testing_methods}

We used three kinds of GW signals to test our constraints: BBH signals observed by the LIGO and Virgo detectors during O1 and O2; simulated BBH signals; and simulated CCSN signals\footnote{Our only intention by applying our methods to these signals was to test and validate our methods. We do not propose to use these constraints to gain new information from signals well described by astrophysical models.}. We analyzed all these signals with the BayesWave algorithm \cite{bayeswave, chirplet}, which robustly reconstructs the waveforms and central moments of signals for a wide range of signal morphologies \cite{BWPE}. We reconstructed all simulated signals from samples of simulated Gaussian noise corresponding to the detectors' design sensitivities.

\begin{table}
\caption{\label{tab:events}Constraints derived for BBH signals GW150914 \cite{O1BBH}, GW151226 \cite{O1BBH}, GW170104 \cite{gw170104}, GW170608 \cite{gw170608}, GW170814 \cite{gw170814} observed by the LIGO-Virgo detector network during the O1 and O2 observing runs. For a detailed description of the constraints, see Section \ref{sec:derivation}. We also show the reference values (medians with 90\% credible intervals) of the parameters. To calculate the lower limits, we used the point estimate of the distance from the model-dependent analysis. Note that our constraints are consistent with the model-based parameter estimates, and they would provide useful information were there no source models available.}
\footnotesize
\begin{indented}
\item[]\begin{tabular}{@{}l|lllll}
\br
 & GW150914 & GW151226 & GW170104 & GW170608 & GW170814\\
\mr
\boldmath $D_{\mathrm{ref}}^{\mathrm{(BBH)}} \ [\mathrm{km}]$ \unboldmath & \boldmath $386^{ + 24 }_{ - 20 }$\unboldmath & \boldmath $129^{ + 35 }_{ - 10 }$\unboldmath & \boldmath $300^{ + 35 }_{ - 30 }$\unboldmath & \boldmath $112^{ + 30 }_{ - 6 }$\unboldmath & \boldmath $330^{ + 20 }_{ - 16 }$\unboldmath\\
$D_{\mathrm{max}} \ [\mathrm{km}]$ & 1563 & 1112 & 1418 & 1457 & 1930\\ 
$D_{\mathrm{min}} \ [\mathrm{km}]$ & 25 & 21 & 33 & 17 & 36\\ 
\mr 
\boldmath $M_{\mathrm{ref}}^{\mathrm{(BBH)}} \ [M_{\odot}]$ \unboldmath & \boldmath $65.3^{ + 4.1 }_{ - 3.4 }$\unboldmath & \boldmath $21.8^{ + 5.9 }_{ - 1.7 }$\unboldmath & \boldmath $50.7^{ + 5.9 }_{ - 5.0 }$\unboldmath & \boldmath $19.0^{ + 5.0 }_{ - 1.0 }$\unboldmath & \boldmath $55.9^{ + 3.4 }_{ - 2.7 }$\unboldmath\\ 
$M_{\mathrm{max}} \ [M_{\odot}]$ & 529.0 & 376.6 & 480.3 & 493.2 & 653.3\\ 
$M_{\mathrm{min}} \ [M_{\odot}]$ & 8.6 & 7.0 & 11.2 & 5.8 & 12.0\\ 
\mr 
\boldmath $d_{L, \mathrm{ref}}^{\mathrm{(BBH)}} \ [\mathrm{Mpc}]$ \unboldmath & \textbf{420}\boldmath $^{ + 150 }_{ - 180 }$\unboldmath & \textbf{440}\boldmath $^{ + 180 }_{ - 190 }$\unboldmath & \textbf{880}\boldmath $^{ + 450 }_{ - 390 }$\unboldmath & \textbf{340}\boldmath $^{ + 140 }_{ - 140 }$\unboldmath & \textbf{540}\boldmath $^{ + 130 }_{ - 210 }$\unboldmath\\ 
$d_{L,\mathrm{max}} \ [\mathrm{Mpc}]$ & 8450 & 21690 & 12860 & 20260 & 13960\\
\br
\end{tabular}\\
\end{indented}
\end{table}
\normalsize

We tested our constraints on five BBH signals detected during O1 and O2: GW150914 \cite{gw150914}, GW151226 \cite{O1BBH}, GW170104 \cite{gw170104}, GW170608 \cite{gw170608}, and GW170814 \cite{gw170814}. We compared our constraints to the reference values of the parameters, which we equate with the point estimates obtained from the model-dependent analysis. We defined the reference value of the characteristic mass as the observed source-frame total mass: $M_{\rm ref}^{\rm (BBH)} \equiv m_1 + m_2$, where $m_1$ and $m_2$ are the observed source-frame component masses. We chose the reference value of the characteristic size to be $D_{\rm ref}^{\rm (BBH)} \equiv 2 (r_1 + r_2)$, where $r_1$ and $r_2$ are the Schwarzschild radii associated with the observed source-frame component masses. Finally, we defined the reference value of the luminosity distance, $d_{L, \mathrm{ref}}^{\rm (BBH)}$, as the luminosity distance estimation from the model-dependent analysis. We listed these reference values of parameters, alongside with the constraints, in Table \ref{tab:events}.

We also tested our constraints on simulated BBH signals, because for these we can directly compare our constraints to the preset parameters of the simulations without relying on results of a model-dependent analysis. We define the reference values of parameters in the same way as for observed BBHs, but using the preset parameter values instead of the observed ones. We used 300 simulated BBH signals for these tests produced with LIGO Algorithm Library's injection infrastructure \cite{lalinj}. We used the \emph{SEOBNRv2threePointFivePN} waveform approximant, which applies the aligned-spin effective-one-body model described in \cite{seobnr}. We set a uniform distribution between 20 $M_{\odot}$ and 50 $M_{\odot}$ for the BBH component masses, and for simplicity, we set the spins of all black holes to zero.

We have also tested our constraints on CCSNe, which are less energetic systems than BBHs, but are among the most promising sources of GW bursts \cite{ccsn_firstgen, sn_bayeswave, ccsn_advanced}. Also, because the GW waveform emitted during a CCSN is non-deterministic \cite{yakunin}, our constraints may help understand the physics of the explosion if we detect such a signal. We used GW waveforms B15-WH07 and B20-WH07 from \cite{yakunin}, which represent simulated GWs emitted from non-rotating 15~$M_{\odot}$ and 20 $M_{\odot}$ CCSN progenitors, respectively. Note that these are 2-dimensional simulations, where only one polarization is available, and the GW amplitude could differ from the one found in 3-dimensional simulations. We compared our characteristic size constraints to the diameter of the iron core right after collapse, which is about 100 km for both waveform simulations (see e.g.~Figure 5 of \cite{yakunin}), i.e.~we set $D_{\rm ref}^{\rm (CCSN)} \equiv 100$ km. We compared our constraints on the characteristic mass to the mass of the iron core, which is about 1.4 $M_{\odot}$ for all massive stars, i.e.~$M_{\rm ref}^{\rm (CCSN)} \equiv 1.4 M_{\odot}$. We scaled the simulated GW signals to make them consistent with a CCSN exploding at a distance of 1 kpc, and we compared our distance constraints to this distance, so $d_{L, \mathrm{ref}}^{\rm (CCSN)} \equiv 1$ kpc.

\subsection{Results}
\label{ssec:results}

\begin{figure}
\centering
    \begin{subfigure}[b]{0.5\textwidth}
            \includegraphics[width=\linewidth]{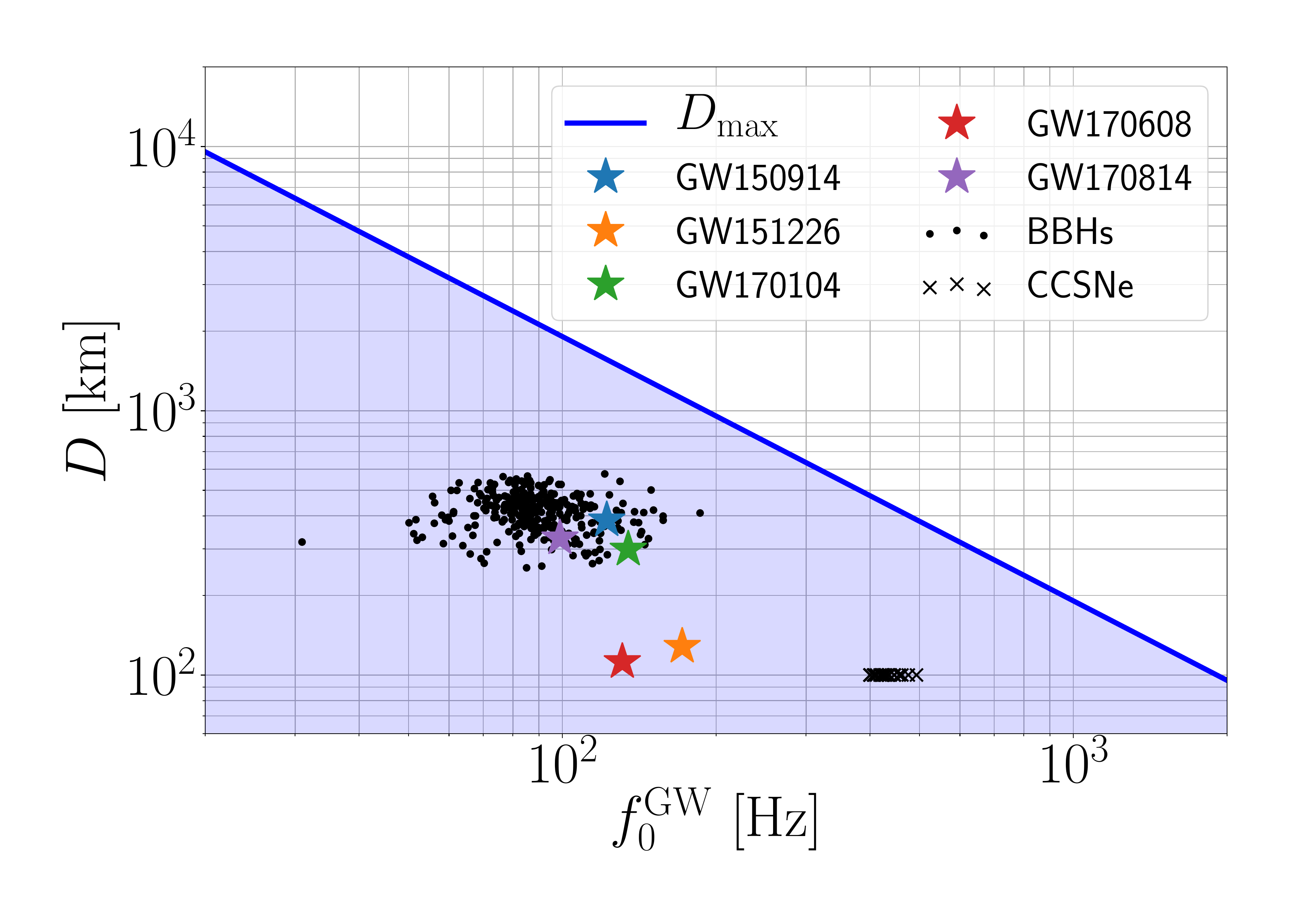}
            \caption{}
            \label{fig:1a}
    \end{subfigure}%
    \begin{subfigure}[b]{0.5\textwidth}
            \includegraphics[width=\linewidth]{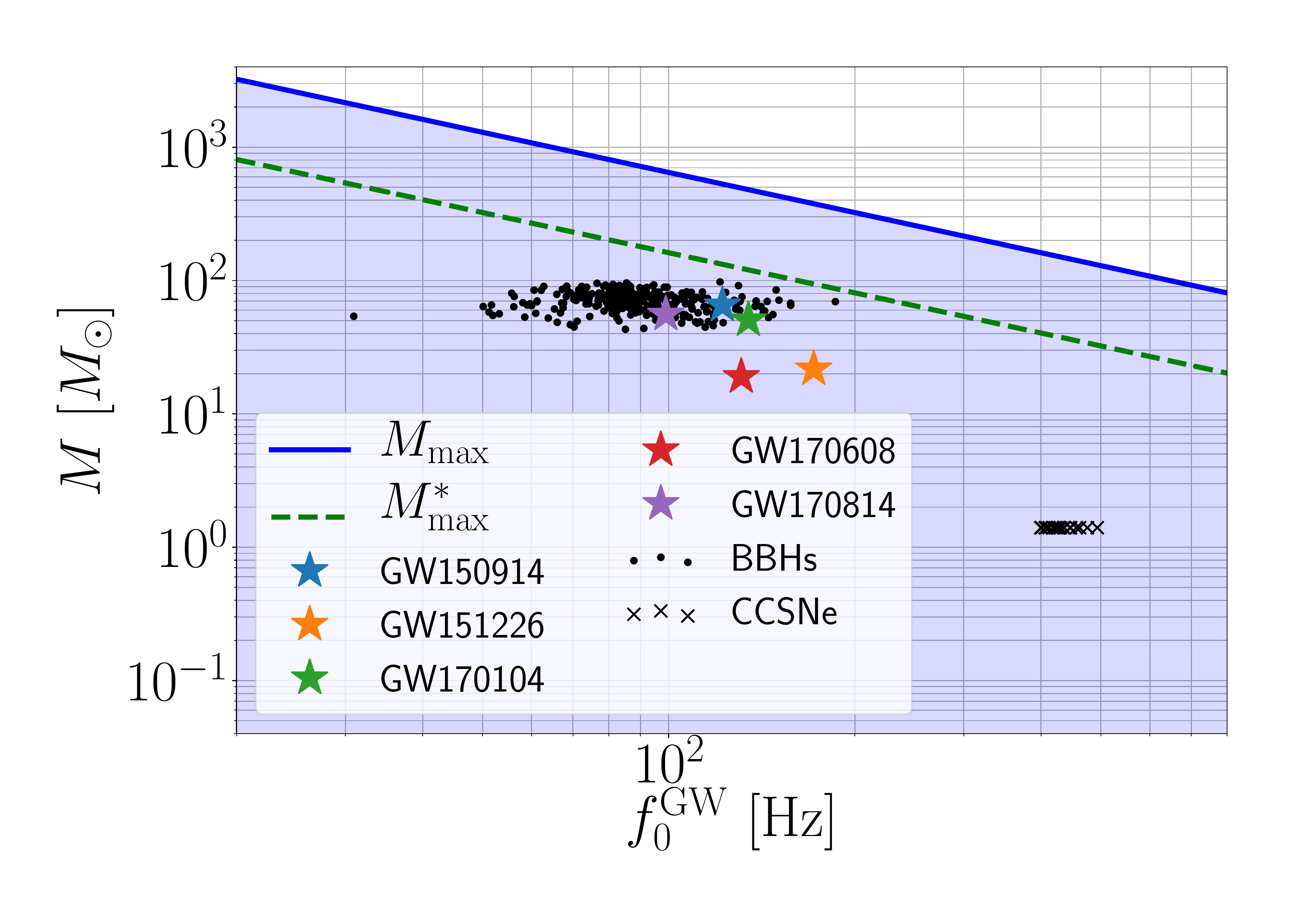}
            \caption{}
            \label{fig:1b}
    \end{subfigure}%

    \bigskip
    \begin{subfigure}[b]{0.5\textwidth}
            \includegraphics[width=\linewidth]{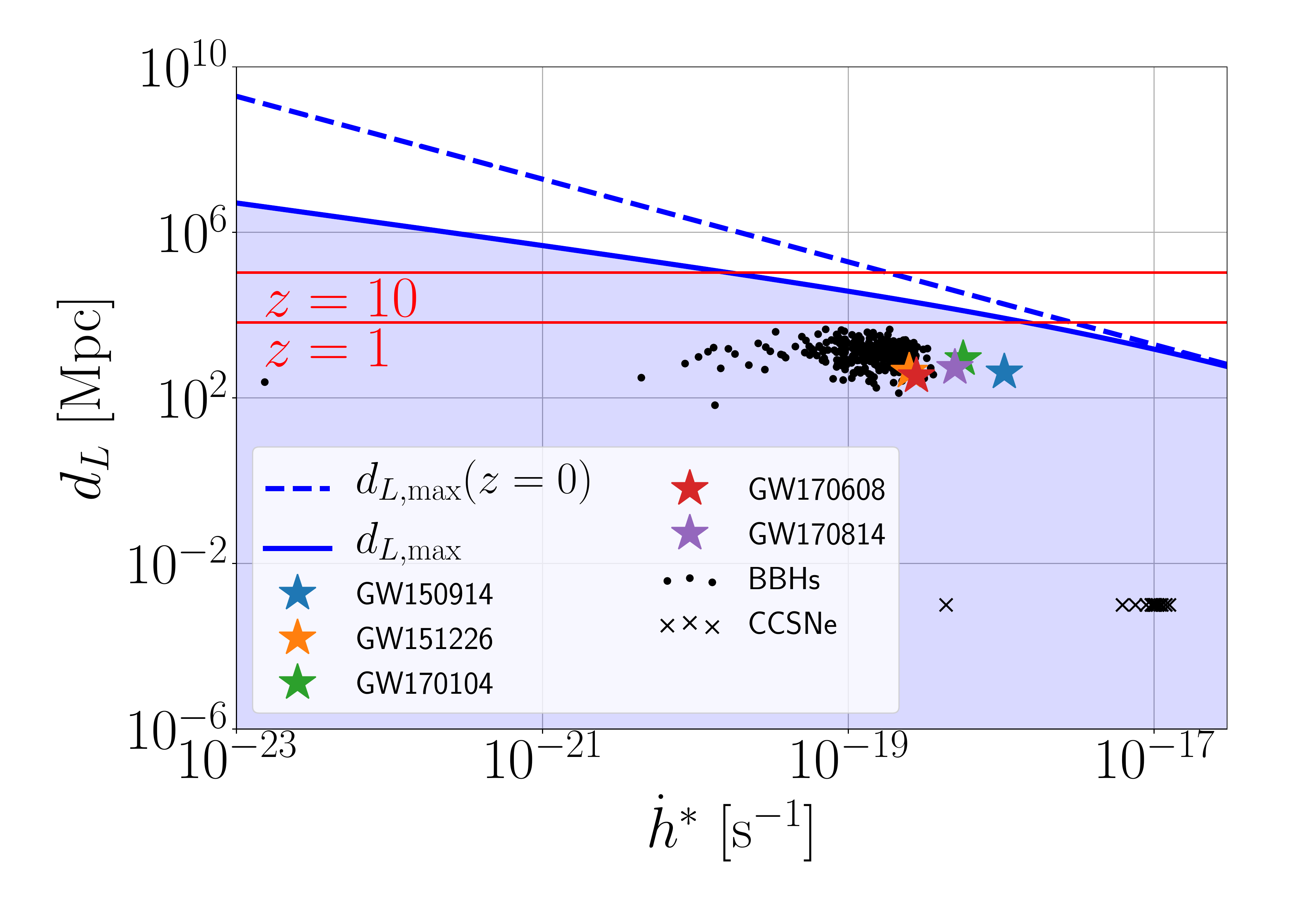}
            \caption{}
            \label{fig:1c}
    \end{subfigure}%
    \caption{Test of upper limits on simulated and observed signals. Blue lines show: (a) upper limit on characteristic size, $D_{\rm max}$, as a function of the reconstructed central frequency ($f_0^{\rm GW}$) of the GW signal; (b) upper limit on characteristic mass, $M_{\rm max}$, as a function of $f_0^{\rm GW}$; and (c) upper limit on luminosity distance, $d_{L, {\rm max}}$, as a function of the reconstructed maximum of the time derivative, $\dot{h}^{*}$, of the GW signal. The blue shaded areas show the regions of the parameter spaces that are consistent with our constraints. We also show the simulated and observed signals with their reference parameter values ($D_{\rm ref}$, $M_{\rm ref}$, $d_{L,{\rm ref}}$) and their relevant model-independent parameter values ($f_0^{\rm GW}$ for panels (a) and (b), and $\dot{h}^{*}$ for panel (c)) reconstructed by BayesWave. Panel (b) also shows $M^*_{\rm max}=0.25 M_{\rm max}$, which is a more restrictive upper limit, but it is only valid for non-spinning equal mass BBHs. On panel (c), we also show the upper limit without redshift correction (see (\ref{eq:zcorr}) in Section \ref{ssec:distance}) with a blue dashed line; and representative distances with horizontal red lines corresponding to $z=1$ and $z=10$. We see that these are all consistent with the upper limits. Observed and simulated BBHs not overlap completely, because we used a very simple mass and distance distribution for the simulated BBHs, which does not describe well the observed BBHs.}\label{fig:upper_limits}
\end{figure}

Figure \ref{fig:upper_limits} show our upper limits along with simulated and observed signals we tested them on. Panel (a) shows $D_{\rm max}$ as a function of $f_0^{\rm GW}$, panel (b) shows $M_{\rm max}$ as a function of $f_0^{\rm GW}$, and panel (c) shows $d_{L, {\rm max}}$ as a function of $\dot{h}^{*}$. The shaded regions represent the parameter spaces consistent with the upper limits. Simulated signals are also plotted with their $D_{\rm ref}$ and $f_0^{\rm GW}$ values for panel (a), $M_{\rm ref}$ and $f_0^{\rm GW}$ values for panel (b), and $d_{L,{\rm ref}}$ and $\dot{h}^{*}$ values for panel (c). $f_0^{\rm GW}$ and $\dot{h}^{*}$ values were reconstructed by BayesWave for all the signals (see Section \ref{ssec:testing_methods} for details). Panel (c) also shows $d_{L, {\rm max}}$ without redshift correction (dashed blue line). We can see that taking the cosmological Doppler-effect into account makes the upper limit significantly more restrictive (especially at low $\dot{h}^{*}$ values). Luminosity distance values corresponding to $z=1$ and $z=10$ are marked with red solid (horizontal) lines on panel (c).

All three panels of Figure \ref{fig:upper_limits} show that all plotted data points fall into the region allowed by our upper limit. This means that our upper limits are valid for all GW signals we used in our tests. This is consistent with our expectations based on the theoretical arguments we presented in Section \ref{sec:derivation}.

%----------------------------------------------------------------------------------------------------------

\begin{figure}
\centering
    \begin{subfigure}[b]{0.5\textwidth}
            \includegraphics[width=\linewidth]{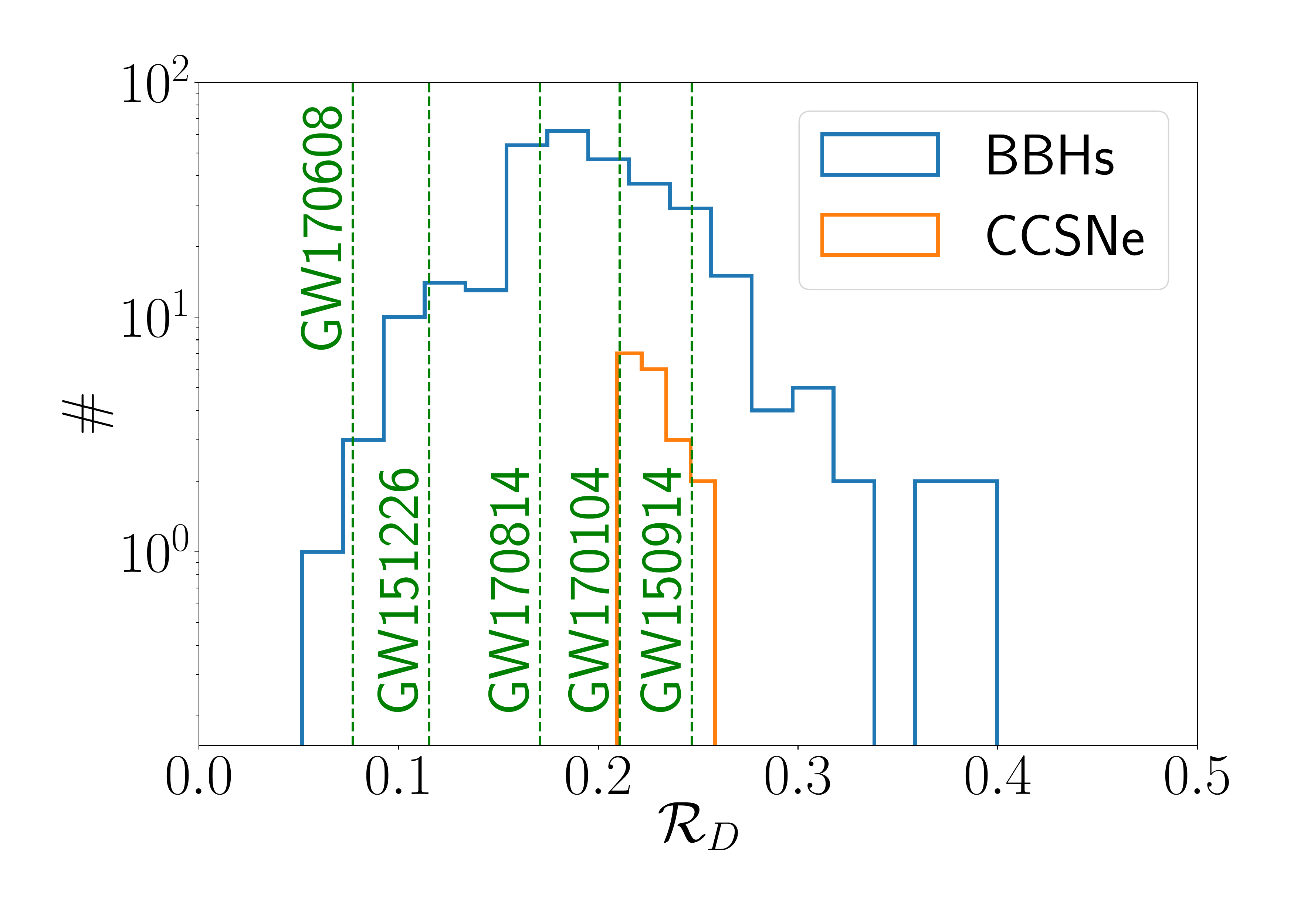}
            \caption{}
            \label{fig:2a}
    \end{subfigure}%
    \begin{subfigure}[b]{0.5\textwidth}
            \includegraphics[width=\linewidth]{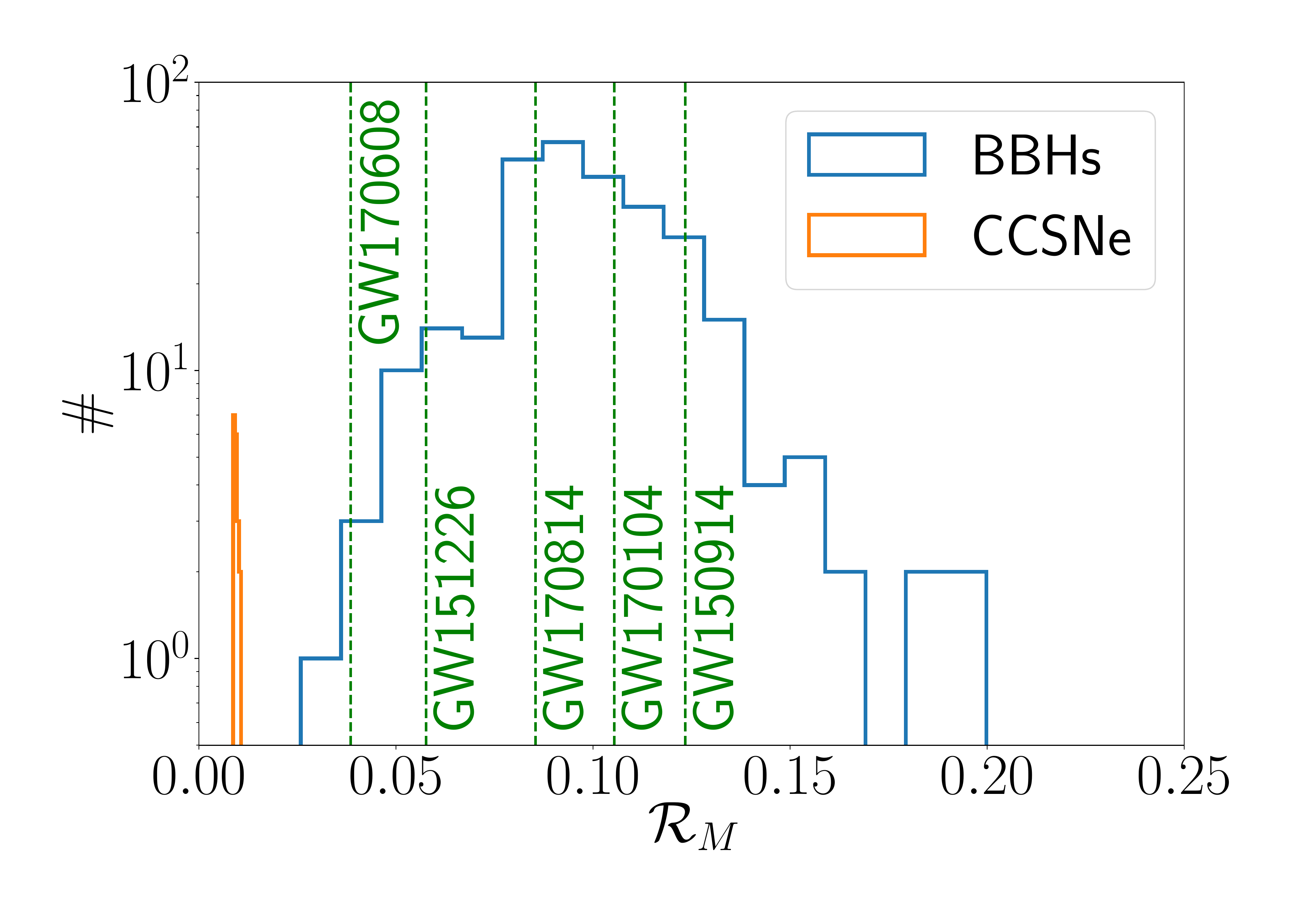}
            \caption{}
            \label{fig:2b}
    \end{subfigure}%

    \bigskip
    \begin{subfigure}[b]{0.5\textwidth}
            \includegraphics[width=\linewidth]{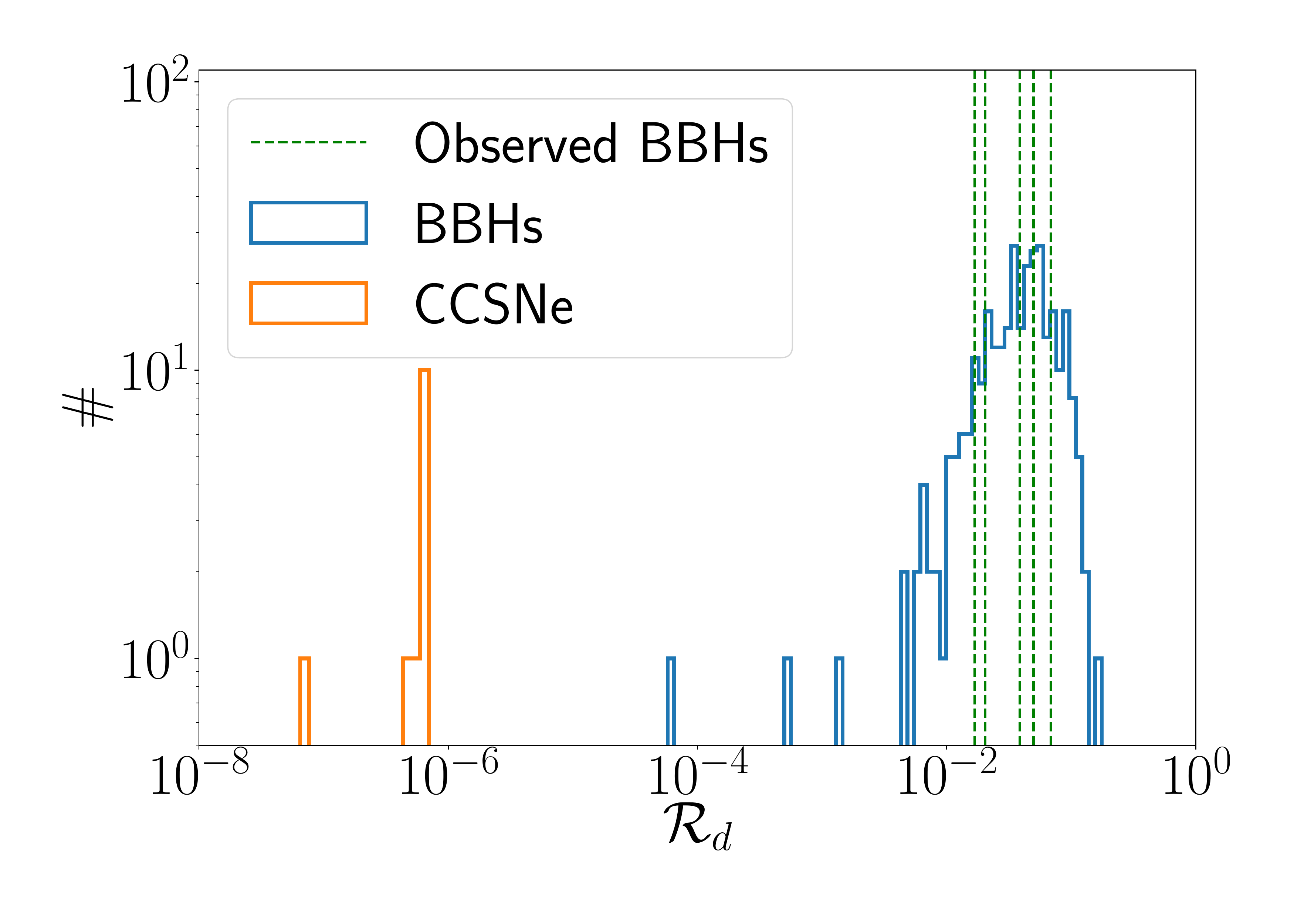}
            \caption{}
            \label{fig:2c}
    \end{subfigure}%
    \caption{Distribution of the ratios of the reference values and the upper limits. Panel (a) shows this ratio for the characteristic size ($\mathcal{R}_D \equiv D_{\rm ref}/D_{\rm max}$), panel (b) shows it for the characteristic mass ($\mathcal{R}_M \equiv M_{\rm ref}/M_{\rm max}$), and panel (c) for the luminosity distance ($\mathcal{R}_{d} \equiv d_{L,{\rm ref}}/d_{L,{\rm max}}$). Blue and orange histograms show the distributions for simulated BBH and CCSN signals, respectively. Observed BBHs are represented individually with green vertical dashed lines. These are individually labeled for panel (a) and (b), while for panel (c) they, from left to right, represent GW170608, GW151226, GW170814, GW150914, and GW170104.}\label{fig:upper_histograms}
\end{figure}

To characterize the efficiency of our upper limits, we compare them to the corresponding reference values. The closer the limit and reference value are to each other, the more we learn about the source from that constraint. Thus we focus on: the $\mathcal{R}_D \equiv D_{\rm ref}/D_{\rm max}$ ratio for the characteristic size limit, the $\mathcal{R}_M \equiv M_{\rm ref}/M_{\rm max}$ ratio for the characteristic mass limit, and the $\mathcal{R}_{d} \equiv d_{L,{\rm ref}}/d_{L,{\rm max}}$ ratio for the luminosity distance limit. These ratios should be below 1 in all cases, and the closer they are to 1, the better our constraints are in characterizing the source. Figure \ref{fig:upper_histograms} show the distributions of these ratios: $\mathcal{R}_D$ in panel (a), $\mathcal{R}_M$ in panel (b), and $\mathcal{R}_{d}$ in panel (c). For simulated signals, we show the histogram of the $\mathcal{R}$ values, while for observed BBHs, we plot their corresponding $\mathcal{R}_D$ values as dashed vertical lines.

From Figure \ref{fig:upper_histograms} (a), it is immediately visible that $\mathcal{R}_D$ is above 5\% for all tested signals, and reaching $\sim$40\% for some of the BBH signals. Also note that the distributions for CCSNe and BBHs are overlapping with each other, which means that our assumptions for the size constraint's limiting case are similarly met for both type of signals. Figure \ref{fig:upper_histograms} (b) shows that $\mathcal{R}_M$ is above 0.8\% and 2.5\% for all simulated CCSN and BBH signals, respectively, and reaching $\sim$20\% for some of the simulated BBH signals. Note that $\mathcal{R}_M$ is significantly lower for CCSN than for BBH signals. This can be understood if we consider that motions in CCSNe are similarly relativistic as in BBHs, but CCSNe are less compact than BBHs (see (\ref{eq:compactness_limit})). Figure \ref{fig:upper_histograms} (c) shows that while the median $\mathcal{R}_{d}$ is 4.0\% for simulated BBHs and 3.9\% for observed BBHs, it is only ($6.7\times10^{-5}$)\% for CCSN signals. So we can see that CCSN signals have $\mathcal{R}_{d}$ ratios orders of magnitudes smaller than BBHs. This difference can be understood as CCSNe are less effective GW emitters than BBHs. While the six orders of magnitude difference between the upper limit and the reference value of distance we see for CCSNe may suggest that the luminosity distance upper limit is not informative, one should keep in mind that for signals coming from an unexpected new source type, these constraints are the only quantitative information we can derive about the source.

%----------------------------------------------------------------------------------------------------------

\begin{figure}
\centering
    \begin{subfigure}[b]{0.5\textwidth}
            \includegraphics[width=\linewidth]{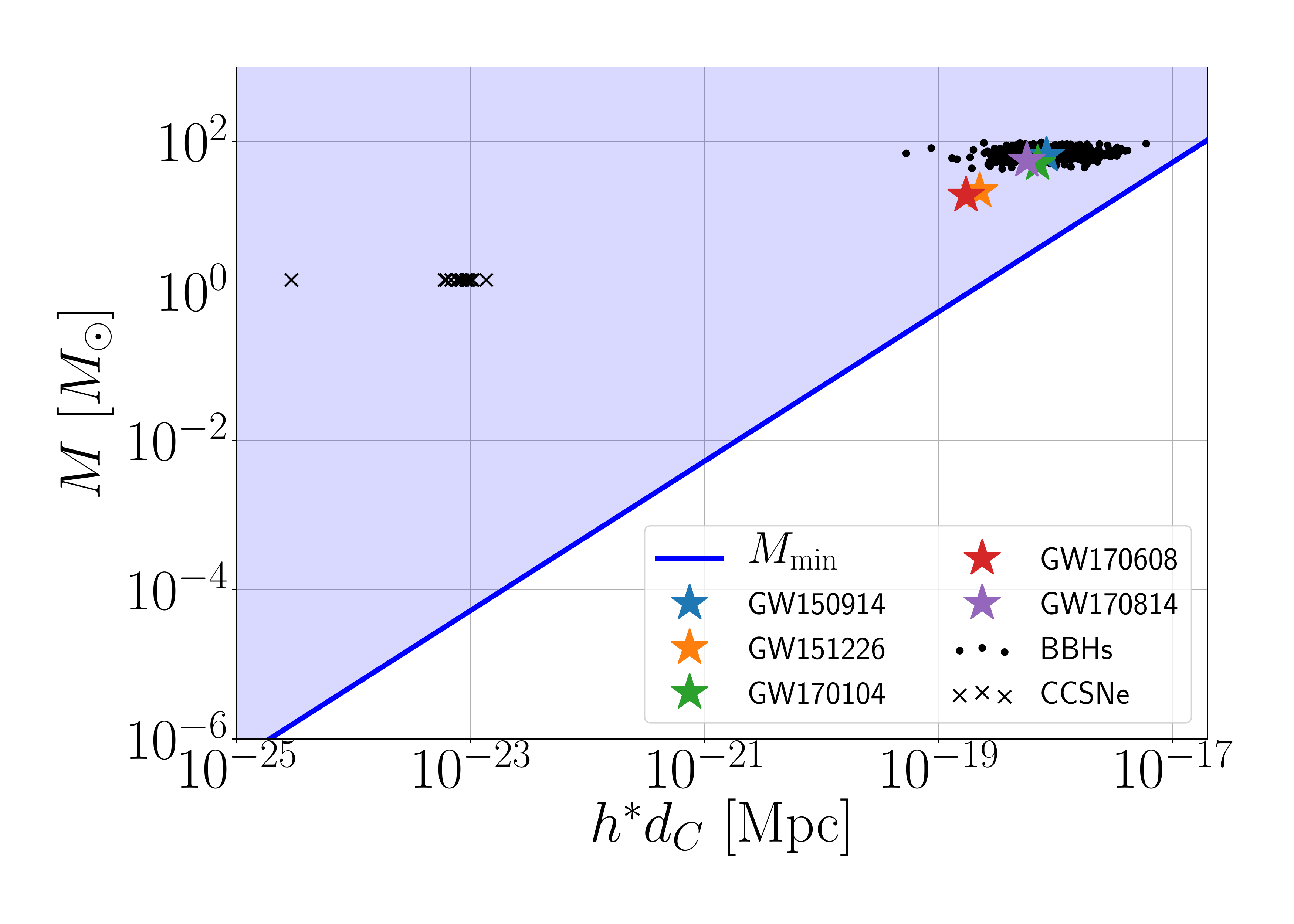}
            \caption{}
            \label{fig:3a}
    \end{subfigure}%
    \begin{subfigure}[b]{0.5\textwidth}
            \includegraphics[width=\linewidth]{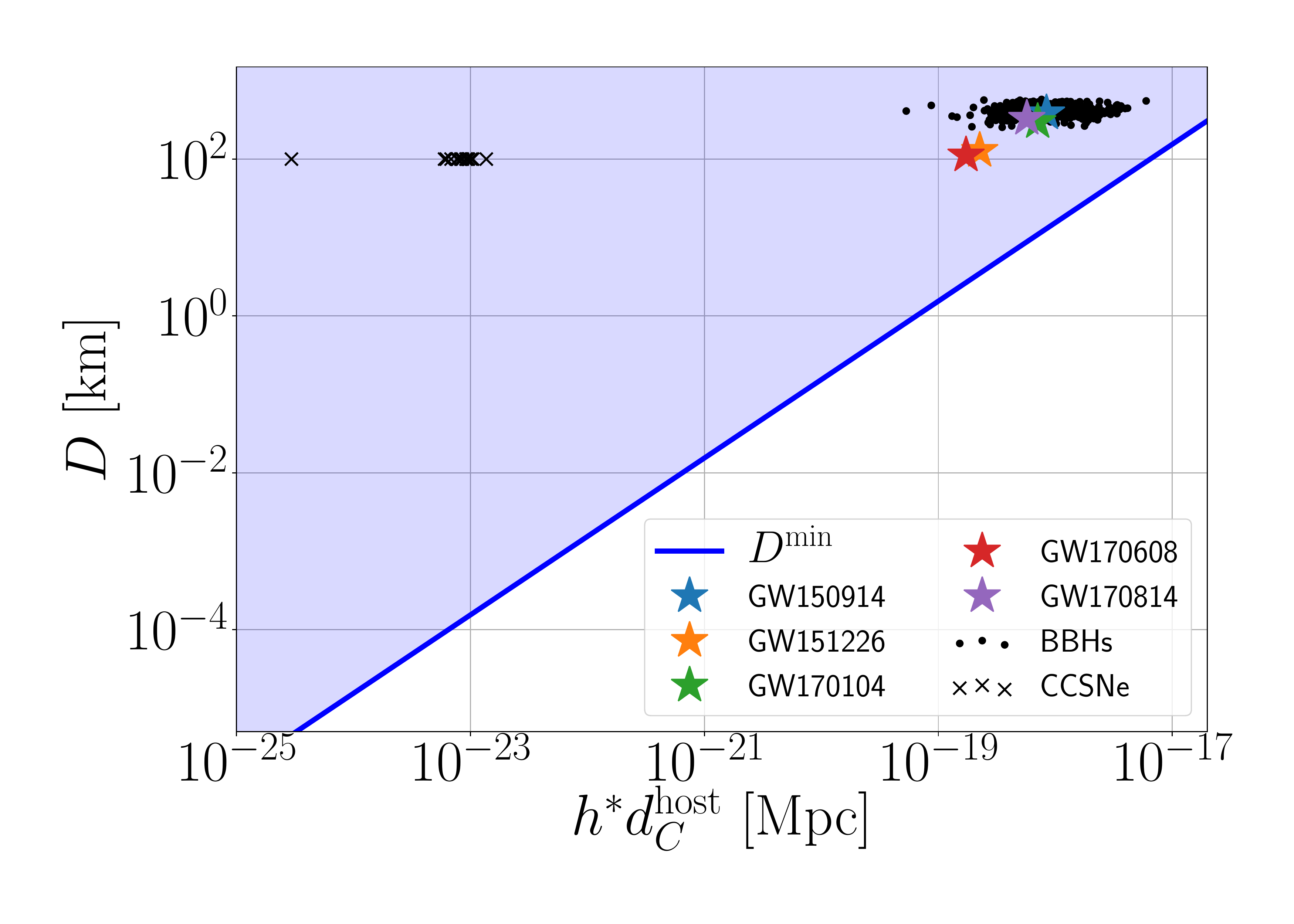}
            \caption{}
            \label{fig:3b}
    \end{subfigure}%
    \caption{Test of lower limits on simulated and observed signals. Blue lines show lower limit on: (a) characteristic mass ( $M_{\rm min}$), and (b) characteristic size ($D_{\rm min}$); both as a function of $h^{*} d_C$, i.e.~the reconstructed maximum of the GW signal times the comoving distance of the source. The blue shaded area shows the region of the parameter space that is consistent with our constraint. We also show the simulated and observed signals with their $M_{\rm ref}$ for panel (a), and $D_{\rm ref}$ for panel (b); and $h^{*} d_C$ values for both panels, where $h^{*}$ was reconstructed by BayesWave. We see that these are all consistent with the lower limit. GW170608 and GW151226 are not overlapping with the simulated BBHs, because we used a 40 $M_{\odot}$ total mass lower cutoff for the simulated BBHs.}\label{fig:lower_limits}
\end{figure}

Figure \ref{fig:lower_limits} shows our lower limits with solid blue lines as a function of $h^{*} d_C$: $M_{\rm min}$ in panel (a), and $D_{\rm min}$ in panel (b). Shaded regions represent the parameter spaces consistent with the lower limits. The simulated and observed signals are also plotted with their $M_{\rm ref}$ or $D_{\rm ref}$ values, and $h^{*} d_C$ values, where $h^{*}$ was reconstructed by BayesWave. All plotted data points fall into the regions allowed by our lower limits, which means that our lower limits are valid for all signals we used in our tests. This is consistent with our expectations based on the theoretical arguments we presented in Section \ref{sec:derivation}.

%----------------------------------------------------------------------------------------------------------

\begin{figure}
\centering
    \begin{subfigure}[b]{0.5\textwidth}
            \includegraphics[width=\linewidth]{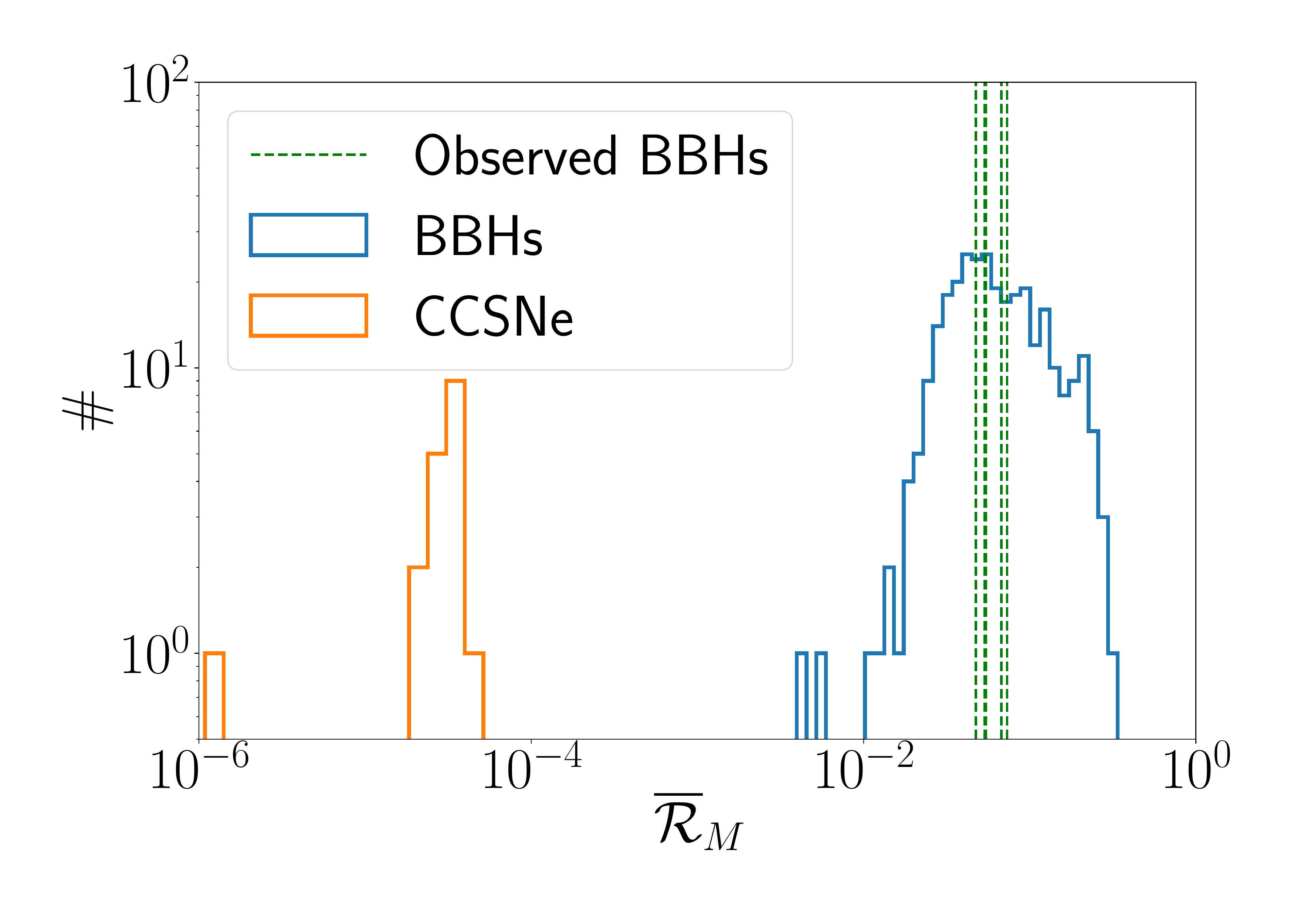}
            \caption{}
            \label{fig:4a}
    \end{subfigure}%
    \begin{subfigure}[b]{0.5\textwidth}
            \includegraphics[width=\linewidth]{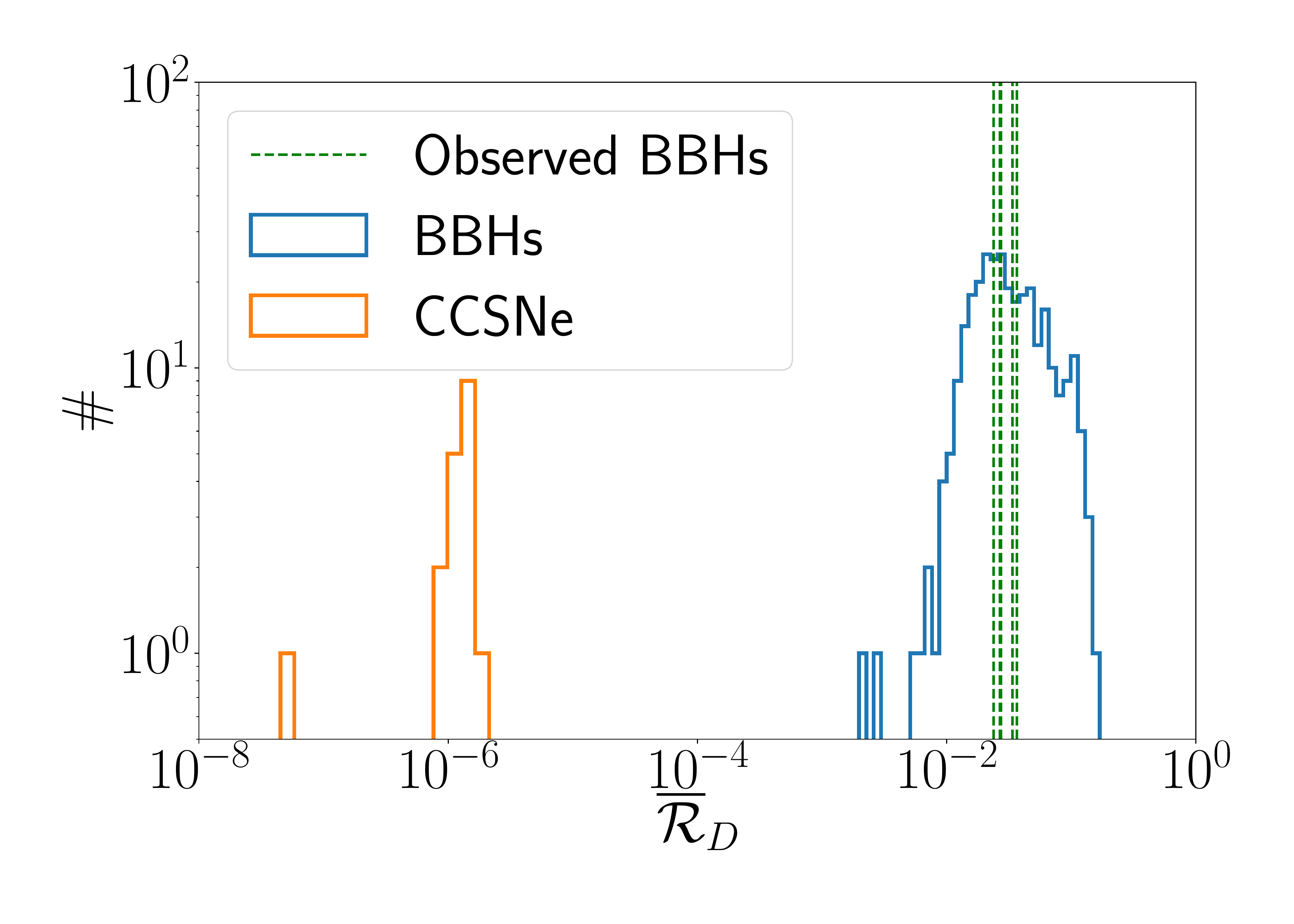}
            \caption{}
            \label{fig:4b}
    \end{subfigure}%
    \caption{Distribution of the ratios of lower limits and reference values. Panel (a) shows this ratio for the characteristic mass ($\overline{\mathcal{R}}_M \equiv M_{\rm min}/M_{\rm ref}$), and panel (b) shows it for the characteristic size ($\overline{\mathcal{R}}_D \equiv D_{\rm min}/D_{\rm ref}$). Blue and orange histograms show the distributions for simulated BBH and CCSN signals, respectively. Observed BBHs are represented individually with green vertical dashed lines.}\label{fig:lower_histograms}
\end{figure}

Similarly to upper limits, to characterize our constraint's efficiency, we should examine the ratio of the lower limit and the reference value. Figure \ref{fig:lower_histograms} (a) shows this ratio for the characteristic mass ($\overline{\mathcal{R}}_M \equiv M_{\rm min}/M_{\rm ref}$), and Figure \ref{fig:lower_histograms} (b) shows it for the characteristic size ($\overline{\mathcal{R}}_D \equiv D_{\rm min}/D_{\rm ref}$). For simulated signals, we show the histogram of $\overline{\mathcal{R}}$ values, while for observed BBHs, we plot their corresponding $\overline{\mathcal{R}}$ values as dashed vertical lines. 

The median value of $\overline{\mathcal{R}}_M$ is 5.8\% for simulated BBHs, 5.4\% for observed BBHs, and ($3.1\times10^{-3}$)\% for CCSNe. The median value of $\overline{\mathcal{R}}_D$ is 2.9\% for simulated BBHs, 2.7\% for observed BBHs, and ($1.3\times10^{-4}$)\% for CCSNe. So we can see that the distributions of $\overline{\mathcal{R}}_M$ and $\overline{\mathcal{R}}_D$ are clearly different for BBHs and CCSNe, with CCSNe being farther away from $\overline{\mathcal{R}}=1$, which can be understood as CCSNe are less energetic GW sources then BBHs.

\section{Discussion}
\label{sec:discussion}

In this section, we address two practical questions to assess the applicability of the proposed constraints. First, we investigate how our constraints would be affected by imperfect waveform reconstruction. Second, we discuss how these constraints can be applied to any GW source, even if they are not naturally described by the parameters we aim to constrain.

\subsection{Effects of imperfect waveform reconstruction}
\label{ssec:wf_reconstruction}

Reliable measurements of signal parameters (which include uncertainties) are essential for giving meaningful constraints on source parameters. These can be obtained from a reconstructed waveform which has corresponding confidence intervals (see \ref{sec:bw_appendix}), i.e.~it covers all the possible waveforms that are consistent with the data at some confidence level. These confidence intervals on the waveform can always be translated to confidence intervals on signal parameters, and thus to confidence intervals on our constraints. In the analysis of an observed signal, one would want to use e.g.~the 90th percentile for upper limits and 10th percentile for lower limits to get constraints at the 90\% confidence level. For simplicity, in this paper we have chosen to use the median values instead. We do not expect this to have a significant effect on our results, because the expected systematic errors are at the 10\% level (see \cite{BWPE}), which is much smaller than the difference between our constraints and the true value of the parameters (see Section \ref{ssec:results}).

Model-independent waveform reconstructions have systematic errors too. Both algorithms and the detectors themselves have varying sensitivities throughout the time-frequency region we analyze, which can lead to systematic inaccuracies in the reconstructions of the waveforms. For example, algorithms providing unmodeled waveform reconstructions (including BayesWave) tend to miss parts of the inspiral phase of low signal-to-noise ratio GWs emitted by low-mass binary black hole systems. Increasing the signal-to-noise ratio would allow the detection of earlier parts of the inspiral, but at any sensitivity, we will always miss the part of the signal where the frequency is outside the sensitive band of the detector. Modeled searches can extract the signal even at frequencies where the detectors are less sensitive by utilizing some waveform model, which we do not have in the scenario examined in this work.

Note that the data can also contain transient non-Gaussian noise features (see e.g.~\cite{gw150914_glitch}), which we have not considered in our analysis. Such a noise transient could bias the estimated parameters of the GW signal if they coincide both in time and frequency \cite{PE_with_glitch}. However, there are ongoing efforts to use the BayesWave algorithm to safely remove such noise transients without compromising the GW signal \cite{bayeswave_glitch_removal, ligo_guide, bayeswave_iii}. Also note that, since glitches tipically occur on the order of once per minute \cite{ligo_guide}, a coincidence is relatively unlikely for short GW bursts with a duration of less than a few seconds.

The sensitivity of our detectors and algorithms is a limitation in reconstructing the GW waveform model-independently, and should be kept in mind when interpreting a detection. This caveat, however, is not specific to our methods described in this paper. It will be present in any analysis that uses the model-independently reconstructed waveform to interpret a signal of unknown origin.

\subsection{Applicability to generic burst sources}
\label{ssec:applicability}

Since we do not assume any source model during their derivation, our constraints can be applied to any GW burst detection, with the caveats that we make a few simplifying assumptions (e.g.~isotropic emission, see details in Section \ref{sec:derivation}). If a particular source family is not naturally described by our constrained parameters ($D$, $M$, and $d_L$), those can be mapped onto the parameters relevant for the given source. Note that we already show an example for such a mapping in Section \ref{ssec:testing_methods}, where we transform the ``characteristic size'' to the Schwarzschild radii of components of a BBH system. Another example is cosmic strings, which are actively being searched for by the LVC \cite{cosmic_string_ligo}, but conventionally are not parameterized with masses. The nominal parametrization uses the string tension ($G \mu$) and invariant length ($l$), or invariant radius ($R$) if a circular geometry is assumed. Nevertheless, $R$ and $G \mu$ can be used to associate a mass (called ``loop mass'') to a string as $M_{\rm loop} = 2 \pi R \mu = l \mu$ (see e.g.~\cite{cosmic_string_loop}), which then can be constrained by our characteristic mass parameter we described in Section \ref{ssec:mass}. This demonstrates that even if a source is not naturally characterized by the same parameters we aim to constrain, our constraints can always be mapped onto the parameter space of the source model.

Note that the constraints presented in this paper can safely be used for GW memory signals too, where the parent signal is not detected (see e.g.~\cite{orphan_memory}). These memory signals will always be detected at lower frequencies compared to their undetected parent burst signals, which means that our mass and size upper limits will give less stringent constraints. \cite{orphan_memory} also showed that the amplitude of the memory signal is always smaller than the parent signal, so our mass and size lower limits will give a smaller (more conservative) constraint. The lower frequency and amplitude also implies that our measured $\dot{h}^{*}$ will be smaller compared to the parent signal, so our luminosity distance upper limit will also give a higher (more conservative) constraint. This analysis shows that if the parent signal does not violate our constraints, the memory signal will not violate it either.

\section{Conclusion}
\label{sec:conclusion}

We suggest, for the first time, methods to characterize transient GW signals using robust constraints we obtain from general relativistic principles, without relying on any astrophysical models of the source. For generic transient GW signals lacking any astrophysical models on the source, such constraints provide the only possibility for a quantitative characterization of the source, and for guiding the development of a source model.

In this paper, our main goal was to provide an approach which can help in interpreting GW bursts originating from new source types. Such an observation can happen at any time with more and more sensitive GW detectors operating around the world. Thus we proposed to use the reconstructed waveform parameters of detected GW signals as inputs for formulae described in Section \ref{sec:derivation} to set upper limits on the characteristic size, characteristic mass and luminosity distance of the source based on the observed GW signal. If information is available on the distance of the source, we can also set lower limits on the characteristic mass and size. Currently this approach provides the only possibility to extract quantitative information about the source of a GW detection without assuming a specific source model and without invoking other (non-GW) observations. Note that our constraints may also be applicable to continuous GW signals (see e.g.~\cite{O2_CW_allsky}), even though we have only tested them for transient ones.

We have tested our constraints on BBHs observed by the LIGO and Virgo GW detectors, as well as on simulated BBH and CCSN signals. We have found that our constraints are always consistent with the preset parameter values of the simulated signals and the point estimates given by the model-based analysis, which means that the proposed constraints are valid for a wide range of sources. The constraints are generally closer to the reference values of parameters for BBH signals than for CCSN signals, and in some cases the reference value of parameters and the constraint are within one order of magnitude. We emphasize that we propose to apply these methods not on GWs from BBHs, but on GW bursts of unknown origin. These efficiency results may not hold for any GW source emitting a GW burst, as we have done our tests using one of the most relativistic sources, where the limiting cases of our constraints are almost reached. However, we need to detect the signal in order to do any parameter estimation, which, considering the sensitivities of current and future GW detectors, already suggests that the source must be highly relativistic, so we expect these results to be representative of the performance of our constraints for wide range of possible GW sources. Also, even in cases when there are several orders of magnitude difference between the reference values of parameters and the constraint, these limits could prove to be useful in interpreting the detection of a generic GW burst without a clear model on its source, because currently this is the only quantitative approach we have for characterizing the source of a GW burst.

Our plan is to incorporate these constraints in the standard output of BayesWave and have them calculated for any transient GW signal observed by the LIGO and Virgo detectors. This way the constraints would be readily available in the case of a detection of a GW burst coming from an unidentified source, and they may help with the interpretation of such a detection. Beyond its straightforward use of informing theorists working on building a model for a new type of GW sources, our constraints could potentially be also used to set prior bounds for particular source models. Based on \cite{BWPE}, we expect that parameter estimation uncertainties will not affect significantly the constraints presented here in the case of a highly significant, high signal-to-noise ratio detection. Calculating statistical uncertainties associated with our constraints are trivial, and planned to be implemented for the actual analysis. We are planning to continue exploring possible ways of interpreting GW bursts. For example, the limits proposed in this paper may be improved by making some simple assumptions on the source. By doing so, we would lose some of the generality but would gain more informative constraints. We are also interested in carrying out a more in-depth analysis on core-collapse supernovae, focusing on how we could better understand the emission mechanism using the constraints presented in this paper. We are also planning to explore the applicability of such constraints for other types of GW detectors, e.g.~LISA \cite{lisa} and Pulsar Timing Arrays (PTAs, see e.g.~\cite{pta_review, pta}).

%The results show that the constraints introduced in Section \ref{ssec:size} for $D$ (see Eq. (\ref{eq:size_limit})), in Section \ref{ssec:mass} for $M$ (see Eq. (\ref{eq:mass_limit})), and in Section \ref{ssec:distance} for $d$ (see Eq. (\ref{eq:distanceupper1}) and (\ref{eq:distanceupper2})) seem to be valid upper limits, i.e.~we did not experienced the violation of them in true detections of binary black hole sources observed during O1. These limits seem to be informative in many cases in the sense that the true value is at the same (or at just one or two lower) order of magnitudes as the limit. Even where this is not true, these upper limits can be useful because for true detections of generic burst signals currently these are the only constraints we can give on these parameters.

\ack
This paper was reviewed by the LIGO Scientific Collaboration under LIGO Document P1800133. We would like to thank Erik Katsavounidis, Jade Powell, and Ronaldas Macas for their valuable comments on the manuscript. Bence B\'ecsy and Peter Raffai was supported by the \'UNKP-17-2 (BB) and \'UNKP-17-4 (PR) New National Excellence Program of the Ministry of Human Capacities of Hungary. The authors would like to acknowledge the use of the LIGO Data Grid computer clusters for performing all the computation reported in the paper. Parts of this research were conducted by the Australian Research Council Centre of Excellence for Gravitational Wave Discovery (OzGrav), through project number CE170100004.

\appendix
\section{Reconstructing model-independent parameters}
\label{sec:bw_appendix}
To calculate the constraints introduced in Section \ref{sec:derivation} we need to estimate three model-independent parameters of the signal: its central frequency ($f_0^{\rm GW}$), its maximum amplitude ($h^{*}$) and the maximum of its time derivative ($\dot{h}^{*}$). These are defined as follows:

\begin{equation}
f_0^{\rm GW} \equiv \int_{0}^{\infty} {\rm d}f \ \frac{2 |\tilde{h}(f)|^2}{h_{\rm rss}^2} f,
\label{eq:f0}
\end{equation}

\begin{equation}
\dot{h}^{*} \equiv \max_{t} \left[ | \dot{h} (t) | \right],
\label{eq:h_dot_max}
\end{equation}

\begin{equation}
h^{*} \equiv \max_{t} \left[ | h(t) | \right],
\label{eq:h_max}
\end{equation}
where $\tilde{h}(f)$ and $h(t)$ are the reconstructed waveform in the frequency domain and in the time domain, respectively, and $h_{\rm rss}$ is the root-sum-squared amplitude defined as ${h_{\rm rss}^2=\int (h_+^2 + h_{\times}^2) \rm{d} t}$.

We can see that all the required parameters can be calculated from the reconstructed waveform. The waveform itself can be extracted from the data in different ways. Here we describe how BayesWave reconstructs the waveform (for more details see e.g.~\cite{bayeswave, BWPE, bayeswave_iii}). BayesWave models the data as Gaussian noise plus a sum of sine-Gaussian wavelets. These wavelets form a frame, to which any time series can be expanded. BayesWave employs a trans-dimensional Markov Chain Monte Carlo sampler, which varies the parameters of the sine-Gaussian wavelets (central time, central frequency, quality factor, amplitude, and phase), as well as the number of wavelets used.  This allows for the accurate reconstruction of GWs with no prior assumption on the signal morphology, but without overfitting the data. As a result, we get a set of wavelets which can be added up for each sample to get a reconstructed waveform at each step of the sampler. These wavelets can be added up for each sample, so we get a reconstructed waveform for each step. From these reconstructed waveforms, we can also compute parameters $f_0^{\rm GW}$, $h^{*}$, and $\dot{h}^{*}$ at each step, from which we can calculate their median values and associated credible intervals. For simplicity, we used the median parameter values in this paper, but an actual analysis should use appropriate percentile values to give constraints at a given confidence level (see Section \ref{ssec:wf_reconstruction} for details).

\section*{References}
%\bibliography{mics_bib}{}
%\bibliographystyle{unsrt_et_al}

\end{document}